Serhii Nazarovets

# Quantitative Analysis of the Co-Publications of Ukrainian Researchers with the 1994—2018 Nobel Laureates in Science

State Scientific and Technical Library of Ukraine, Kyiv, Ukraine
serhii.nazarovets@gmail.com
http://orcid.org/0000-0002-5067-4498

***Introduction.*** The Nobel Prize is awarded annually for outstanding scientific discoveries and inventions. Most scientific papers today are co-authored by a large number of researchers. However, very few scientists can receive the Nobel Prize according to the Statutes of the Nobel Foundation. An analysis of the co-authorship of the Nobel laureates will make it possible to identify employees of Ukrainian institutions who have collaborated with leading scientists of the world, whose scientific works were noted by Nobel.

***Problem Statement.*** For the development of science in Ukraine it is important to study the successful experience of cooperation of domestic research institutions with leading world scientists and research centers, because international scientific collaboration facilitates the process of acquiring new knowledge, promotes mutual enrichment of ideas, efficient use of resources and expands opportunities for further use of research results.

***Purpose.*** Explore the network of collaborators research institutions of Ukraine workers with Nobel laureates in selected scientific fields.

***Materials and Methods.*** Created a list with Scopus ID Nobel laureates 1994—2018 in the fields of Physics, Chemistry, Medicine or Physiology. Using the Scopus database, selected publications of Nobel Prize winners, which were written in collaboration with scientists who worked in Ukrainian institutions. The number of these publications, their authors, the type, time of writing and the number of citations were determined.

***Results.*** The 31 publications were singled out, in which the Nobel laureates of 1994—2018 and employees of scientific institutions of Ukraine were co-authors. A total of 37 such authors from 14 scientific institutions have been identified.

***Conclusions.*** The data obtained indicate that the employees of scientific institutions of Ukraine published very few papers in collaborations with Nobel Prize winners of 1994—2018 in comparison with employees of institutions in leading countries in publishing activity. Consequently, the system of relations of Ukrainian institutions with foreign scientific institutions, whose employees make an important contribution to scientific progress, is underdeveloped.

*Keywords:* Nobel Prize, Nobel Prize laureates, employees of Ukrainian institutions, co-authors, quantitative analysis.

The Nobel Prize is one of the most prestigious and well-known international prizes, which is awarded annually for significant scientific discoveries or inventions. According to the statute of the Nobel Foundation, which regulates the rules for awarding the prize, one or two works can be awarded at the same time, and the total number of winners shall not exceed three people [1]. Since the fact of winning the Nobel Prize can be considered a certain indicator of the scientific achievements of researcher, there have been many scientometrics studies of the publishing activity of its winners.

Nobel Prize winners tend to be more productive, as they start publishing scholarly research papers earlier and more often than other researchers and, consequently, publish more research works during their careers [2]. The study of the publishing activity of Nobel prize winners in the field of chemistry and medicine has shown that they, on the contrary, have fewer publications than researchers in other fields, but they publish more individual works, both before and after receiving the honorary award [3]. As the research works of Nobel Prize winners contribute to the emergence of new original scientific ideas, as a rule, their publications are cited much more often than the publications of average researchers. Based on data on the number of citations of scholarly research publications, Eugene Garfield has identified highly cited researchers as "Nobel level" researchers, i.e. those who are likely to receive the Nobel Prize [4].

However, to predict future Nobel Prize winners, it is not enough to have only data on the citation of research works. Many other factors shall be taken into account. In addition, it has been found that the publications of researchers after receiving the Nobel Prize are cited much more often than before receiving the award, which is fully consistent with the "Matthew effect in science" [5, 235]. Also, it has been also proved that the fact of receiving the Nobel Prize triggers a sort of "chain reaction" of citations, because after receiving the award, the number of citations of not only the awardee publications, but also those mentioned in his/her research increases [6].

Many studies have established a relationship between age and productivity of Nobel Prize winners in various fields of science [7—10]. It has been shown that Nobel laureates with theoretical education receive more awards than those who deal with empirical research [11]; the female prize winners in physics, chemistry, and medicine more seldom get married and have fewer children and publications as compared with the male prize winners [12]. In addition, with the help of publications by Nobel Prize winners the suitability of using the Google Scholar database for scientometrics research has been tested [13], and certain scientometrics indicators have been verified [14].

Studies of institutions and countries in which Nobel Prize winners conducted research in physics, chemistry and medicine or physiology have shown the leadership of the United States. Most nominees change jobs either having received a PhD or having published the papers that later became decisive for the Nobel Prize. However, the researchers move mostly to institutions in the same country. It has been found that having been awarded; the Nobel Prize winners change their approach to scientific cooperation, as the number of works written by them with new co-authors decreases [15].

No researcher of a R&D institution in Ukraine has yet been awarded the Nobel Prize. However, it may be assumed that they have worked together and shared publications with Nobel Prize winners, and thus contributed to the Nobel Prize-winning discoveries or, at least, have a well-developed network of international scientific contacts that allow them to collaborate with leading researchers from around the world in award-winning scientific fields.

Until now, the co-authorship in professional publications of researchers of R&D institutions of Ukraine with Nobel Prize winners in the fields of physics, chemistry, and medicine or physiology has not been studied. Therefore, the questions to be answered are as follows:

1. How many publications have Ukrainian researchers published together with Nobel laureates and how significant is this number for each scientific field?

2. In which R&D institutions do or did the authors of joint publications with Nobel laureate work?

From the Scopus database with the help of tool for tracking scholarly research citations, the publications of Nobel laureates of 1994—2018 in physics, chemistry, medicine or physiology have been analyzed, and the research works published in co-authorship with researchers who indicate that they are employed in Ukrainian institutions have been selected.

To do this, at the first stage of the study, a list of Nobel laureates in 1994—2018 in physics, chemistry, medicine or physiology has been compiled according to the information presented on the Nobel Prize official website (https://www.nobelprize.org/prizes/), and the Scopus ID of each winner has been counted. This list is freely available at Zenodo for further research (http://doi.org/10.5281/zenodo.3242024).

At the second stage of the study, on April 19, 2019, based on the data from the Scopus database, the Nobel Prize winner publications co-authored with researchers who indicated that they were employed in Ukrainian institutions were listed. For example, the search query for the 2018 winners looked like this: AU-ID ("Ashkin, Arthur A." 7003716134) OR AU-ID ("Mourou, Gérard A." 7102620818) OR AU-ID ("Strickland, Donna T. "56277784300) AND (LIMIT-TO (AFFILCOUNTRY," Ukraine ")). In addition, the information on the year, institution, city, and country listed in Scopus was verified manually. In the case of errors, the research papers are not taken into account. For example, in the case of publication https://doi.org/10.1016/0038-1098(92)90174-8, where Ukrainian address was erroneously indicated for the institution of the Russian Federation, or with publications https://doi.org/10.1070/PU1987v030n02ABEH002815 and https://doi.org/10.1070/PU1987v030n07-ABEH002932 published in the Soviet times).

For each joint publication of researchers from Ukrainian institutions with 1994—2018 Nobel laureates in physics, chemistry, medicine or physiology, the number of co-authors and the number of citations of this publication have been determined. Also, Field-Weighted Citation Impact (FWCI) has been defined for each publication. It shows how often a particular work is cited as compared with similar works of the same type, year, and industry [16]. In addition, joint publications of researchers from institutions of Ukraine's neighboring countries with the 1994—2018 Nobel laureates have been counted.

The works of Ukrainian researchers affiliated with foreign institutions, as well as the works of researchers of Ukrainian origin (such as Nobel laureates Roald Hoffman or Georges Sharpack) are not taken into consideration. At the same time, the works of researchers of different nationalities who work in institutions of Ukraine are taken into consideration. It should be noted that some joint publications of Ukrainian researchers with Nobel laureates could be published in journals not represented in Scopus, or their archives is not fully represented in this database, or the institutional affiliation of the authors is missing in Scopus (Undefined) records, therefore, the data may be incomplete.

The researchers who indicate in publications that they work in R&D institutions of Ukraine, have published in co-authorship with the 1994—2018 Nobel laureates 31 publications (various types of publications, except for Erratum): 25 in physics (21 articles, 3 conference proceedings, and 1 review), 4 in chemistry (4 articles), and 2 in medicine or physiology (1 review and 1 letter).

Globally, the largest number of publications co-authored with the 1994—2018 Nobel Laureates in physics, chemistry, and medicine or physiology has researchers from U.S. institutions. In Europe, the undisputed leaders are researchers of institutions in the United Kingdom, Germany, and France (https://doi.org/10.5281/zenodo.3900998). Ukrainian researchers have published very few joint publications with Nobel laureates, even as compared with researchers of institutions in most of Ukraine's neighboring countries (Table 1).

*Table 1.* **Joint Publications of Researchers from R&D Institutions of Ukraine's Neighboring Countries with the 1994–2018 Nobel Prize Winners, number of publications**

| Country | Field of science | | |
| --- | --- | --- | --- |
| | Physics | Chemistry | Medicine or physiology |
| Russia | 1774 | 84 | 26 |
| Poland | 403 | 62 | 26 |
| Hungary | 177 | 44 | 26 |
| Romania | 69 | 5 | 2 |
| Ukraine | 25 | 4 | 2 |
| Belarus | 5 | 1 | 0 |
| Slovakia | 1 | 8 | 8 |
| Moldova | 1 | 0 | 1 |

Fifteen Ukrainian institutions whose employees worked together with the 1994—2018 Nobel laureates have been identified. Most of them are institutions of the National Academy of Sciences of Ukraine (Table 2). Also, 37 authors who at the time of publishing worked in Ukrainian R&D institutions and have joint publications with 16 Nobel laureates in 1994—2018 in physics, chemistry, medicine or physiology have been identified. Table 3 shows researchers who have, at least, two joint publications.

All these publications are co-authored by several researchers, sometimes the co-authors work together at one institution or in different institutions of Ukraine. Therefore, in Tables 2 and 3, the same publication may be credited to several institutions or researchers at once. For example, three works of the Verkin Physical and Technical Institute for Low Temperatures of the NAS of Ukraine are co-authored by three employees of this institute, and the research https://doi.org/10.1002/ijc.25414 is published in collaboration with employees of the two Ukrainian institutions: RE Kavetsky Institute of Experimental Pathology, Oncology, and Radiobiology of the NAS of Ukraine and Lashkariov Institute for Physics of Semiconductors of the NAS of Ukraine.

Almost half of the total number of joint research works by researchers of Ukrainian institutions in collaboration with the 1994—2018 Nobel laureates is in the field of physics (12 publications), published by teams of more than 10 co-authors: 3 publications are made by 8 co-authors, 2 publications by 7 co-authors, 2 publications by 6 co-authors, 2 publications by 5 co-authors, 3 publications by 4 co-authors, and 1 publication by 3 co-authors. It has been found that 4 joint publications in the field of chemistry have 5, 6, 8, and 10 co-authors, and both researches in the field of medicine or physiology have more than 10 co-authors.

*Table 2.* **R&D Institutions of Ukraine Whose Researchers have Co-Publications with the 1994—2018 Nobel Prize Winners**

| Institution | Subordination | Number of co-publications | Type of publication |
|---|---|---|---|
| Frantsevich Institute for Problems of Materials Science of the NAS of Ukraine | The NAS of Ukraine | 6 | 3 articles, 2 letters, 1 conference proceedings |
| Verkin Physico-Technical Institute for Low Temperatures of the NAS of Ukraine | The NAS of Ukraine | 5 | 4 articles, 1 conference proceedings |
| Donbas State Machine-Building Academy | The MES of Ukraine | 3 | 3 articles |
| Crimean Astrophysical Observatory | The MES of Ukraine | 3 | 3 articles |
| Galkin Physico-Technical Institute of the NAS of Ukraine | The NAS of Ukraine | 2 | 1 стаття, 1 letter |
| Institute of Physics of the NAS of Ukraine | The NAS of Ukraine | 2 | 2 articles |
| Lashkariov Institute for Physics of Semiconductors of the NAS of Ukraine | The NAS of Ukraine | 2 | 2 articles |
| Institute for Cell Biology and Genetic Engineering of the NAS of Ukraine | The NAS of Ukraine | 2 | 1 article; 1 review |
| Sukhomlynskyi National University of Mykolaiv | The MES of Ukraine | 2 | 2 articles |
| RE Kavetsky Institute for Experimental Pathology, Oncology, and Radiobiology of the NAS of Ukraine | The NAS of Ukraine | 1 | Article |
| Bogomolets Institute of Physiology of the NAS of Ukraine | The NAS of Ukraine | 1 | Review |
| Taras Shevchenko National University of Kyiv | The MES of Ukraine | 1 | Article |
| Institute of Single Crystals, R&D Complex of the NAS of Ukraine | The NAS of Ukraine | 1 | Review |
| Kharkiv Physico-Technical Institute, National R&D Center | The NAS of Ukraine | 1 | Conference proceedings |
| Sumy State University | The MES of Ukraine | 1 | Article |

Table 3. Researchers Who Have, At Least, Two Co-Publications with 1994—2018 Nobel Prize Winners

| Name | Employment | Number of publications | Name | Employment | Number of publications |
|---|---|---|---|---|---|
| Z.D. Kovaliuk | Frantsevich Institute for Problems of Materials Science of the NAS of Ukraine | 4 | A.A. Sybirnyi | Institute for Cell Biology and Genetic Engineering of the NAS of Ukraine | 2 |
| M.M. Gorkavyi | Crimean Astrophysical Observatory | 3 | V.N. Tulupenko | Donbas State Machine-Building Academy | 2 |
| N.V. Krainiukova | Verkin Physico-Technical Institute for Low Temperatures of the NAS of Ukraine | 3 | S.A. Uiutnov | Verkin Physico-Technical Institute for Low Temperatures of the NAS of Ukraine | 2 |
| S.S. Guziy | Sukhomlynskyi National University of Mykolaiv | 2 | O.E. Filippov | Galkin Physico-Technical Institute of the NAS of Ukraine | 2 |
| V.K. Dugaiev | Frantsevich Institute for Problems of Materials Science of the NAS of Ukraine | 2 | I.V. Khyzhnyi | Verkin Physico-Technical Institute for Low Temperatures of the NAS of Ukraine | 2 |
| O.V. Savchenko | Verkin Physico-Technical Institute for Low Temperatures of the NAS of Ukraine | 2 | | | |

The number of joint research works of employees of Ukrainian institutions with 1994—2018 Nobel laureates, which are published before and after receiving the Nobel Prize, is divided almost equally: 18 and 17 publications, respectively (in general, there are more joint works because of the fact that co-authors of several works are laureates of different years).

The Field-Weighted Citation Impact (FWCI) metric is available for publications in the Scopus database and shows how often other researchers have mentioned a particular work in their publications as compared with similar works published in the same year and in the same field. If this indicator is equal to 1, it means that the work is cited as expected, if the value of the indicator is > 1, the work is cited better than expected, and if it is <1, this may indicate that the research is insufficiently influential. Because this FWCI takes into account the field of research, chronology, and type of publication, it is considered a better indicator of assessing the impact of research than a mere count of the number of citations [16]. According to the FWCI, almost half of the joint publications of Ukrainian researchers with the Nobel laureates in physics are cited above the world average; in the field of chemistry, all collaborations have a score much lower than the world average, and in medicine or physiology the index is much better than the world average (https://doi.org/10.5281/zenodo.3901000).

The study of the publishing activity of researchers from R&D institutions of Ukraine in co-authorship with the 1994—2018 Nobel laureates in physics, chemistry, medicine or physiology, has shown that over the past quarter of the century, Ukrainian researchers have published very few such research works as compared with not only researchers from the institutions that are leaders in scholarly research publication, but also their colleagues from Ukraine's neighboring countries. Scientific cooperation is not limited to co-authorship [17], but such a small number of joint research works does not allow stating that R&D institutes and universities of Ukraine have a developed system of relations with foreign research institutions whose researchers make important contributions to outstanding scientific discoveries and inventions for which the Nobel Prize is awarded.

This paper does not consider possible explanations for low joint publishing activity of researchers from Ukrainian institutions with Nobel laureates, however, it is likely that the primary reason is limited funding for research institutions, as Nobel-class research requires access to advanced tools and materials [18]. At the same time, several researchers from Ukrainian institutions have co-publications with Nobel laureates, and this experience of cooperation shall be studied and taken into account in the development of an effective national plan of activities of educational and research institutions.

International co-authorship usually leads to publishing research papers with higher citation indexes, as shown in previous studies [19]. The vast majority of the frequently cited publications analyzed in this paper (according to FWCI) have more than 10 co-authors from different countries. This is especially noticeable in the case of analyzed publications in chemistry and medicine or physiology, where all considered works in the field of chemistry have less than 10 co-authors and FWCI <1, while in the field of medicine or physiology the publications have more than 10 co-authors and FWCI> 11, respectively.

While developing their scientific careers, researchers may several times change their institutional affiliation, so it would be worthwhile to examine how the network of co-authors is affected by the change of affiliation, in particular, researchers who previously worked or start working in institutions of Ukraine. This may be a subject of the further research.

Decisions to award scientific prizes, including the Nobel Prize, can be influenced by various factors that are not directly related to the study itself, and not all researchers who have been awarded are able to receive them for various reasons [20]. Therefore, the results of this study shall not be used in the context of evaluating the activities of Ukrainian R&D institutions. However, the data presented in this paper may be useful to draw the attention of officials and the public to the problem of international scientific cooperation of Ukrainian institutions.